\documentclass[a4paper,longauth]{aa}
\usepackage{txfonts}
\usepackage{graphicx,subfigure}
\usepackage{hhline}
\usepackage{epsfig}
\usepackage{epsf}
\usepackage{lscape}
\usepackage[square]{natbib}
\usepackage{rotating}
\usepackage{tabularx}
\usepackage{lscape}
\usepackage{supertabular}
\usepackage{multirow}
\usepackage{wasysym}
\usepackage{hyperref}
\newcommand{\mic}{\,{\rm \mu m} }
\newcommand{\be}{\begin{equation}}
\newcommand{\ee}{\end{equation}}
\newcommand{\bea}{\begin{eqnarray}}
\newcommand{\eea}{\end{eqnarray}}
\newcommand{\ba}{\begin{array}}
\newcommand{\ea}{\end{array}}

\begin{document}   
\title{An analysis of star formation with Herschel in the Hi-GAL Survey.
I. The Science Demonstration Phase Fields.}

\author{M. Veneziani\inst{1}, D. Elia\inst{2}, A. Noriega-Crespo\inst{1}, R. Paladini\inst{1}, S. Carey\inst{1}, A. Faimali\inst{3}, S. Molinari\inst{2}, M. Pestalozzi\inst{2}, F. Piacentini\inst{4}, E. Schisano\inst{2}, C. Tibbs\inst{1}} 
 
\institute{Infrared Processing and Analysis Center, California Institute of Technology, Pasadena, CA, 91125
\and
INAF-IFSI - Via Fosso del Cavaliere 100, Rome, Italy
\and
School of Physics, Astronomy and Mathematics, University of Hertfordshire, College Lane, HatÞeld, AL10 9AB, UK 
\and
Dipartimento di Fisica, Universita di Roma ``La Sapienza", P.le Aldo Moro 2, 00185, Rome, Italy 
}

\offprints{marcella.veneziani@ipac.caltech.edu}

\abstract
{}
{ The Herschel survey of the Galactic Plane (Hi-GAL) provides a unique
opportunity to study star formation over large areas of the sky and
different
environments in the Milky Way. We use the best studied Hi-GAL fields to date, two $2^\circ\cdot2^\circ$ tiles centered on ($\ell$, $b$) = ($30^\circ,\;0^\circ$) and ($\ell$, $b$) = ($59^\circ,\;0^\circ$), to study the star formation activity in these regions of the sky using a large sample of well selected
young stellar objects (YSOs).}
{ We use the Science Demonstration Phase Hi-GAL fields, where a
tremendous effort
has been made to identify the newly formed stars and to derive as accurately as
possible their properties such as distance, bolometric luminosity,
envelope mass and stage of evolution. We estimate the star formation
rate (SFR) for
these fields using the number of { candidate} YSOs and their average time scale to
reach the Zero Age Main Sequence, and compare it with the rate estimated
using their
integrated luminosity at 70~$\mu$m combined with an extragalactic star formation
indicator.}
{We measure a SFR of ${(9.5\pm4.3)\cdot10^{-4}\;{\rm M_\odot/yr}}$ and ${(1.6\pm0.7)\cdot10^{-4}\;{\rm M_\odot/yr}}$ with the source counting method, in $\ell=30^\circ$ and $\ell=59^\circ$, respectively. 
Results with the 70~$\mu$m estimator are ${(2.4\pm0.4)\cdot10^{-4}\;{\rm M_\odot/yr}}$ and ${(2.6\pm1.1)\cdot10^{-6}\;{\rm M_\odot/yr}}$. Since the 70~$\mu$m indicator is derived from averaging extragalactic star forming complexes, we perform an extrapolation of these values to the whole Milky Way and obtain SFR$_{MW}{ = (0.71\pm0.13)\;{\rm M_\odot/yr}}$ from l=30$^\circ$ and SFR$_{MW}{ = (0.10\pm0.04)\;{\rm M_\odot/yr}}$ from $\ell=59^\circ$. The estimates in $\ell=30^\circ$ are in agreement with the most recent results on the Galactic star formation activity.  }
{{ The source counting method gives results valid only for the particular region under consideration. On the contrary, the IR indicator, by construction,  gives 
results which can be extrapolated to the whole Galaxy. In particular, when the extragalactic indicator is applied to the $\ell=30^\circ$ field, it provides a SFR consistent with previous estimates, indicating that the characteristics of this field are likely close to those of the star-formation dominated galaxies used for its derivation.}
Since the sky coverage is limited, this analysis will improve when the full Hi-GAL survey will be available. It will cover the whole Galactic Plane, sampling almost the totality of Galactic star forming complexes. By means of the { candidate YSOs} counting method it will be then possible to calibrate a SFR Galactic indicator and to test the validity of the extragalactic estimators.  }


\authorrunning{M. Veneziani et al.}
\titlerunning{SFR in the Hi-GAL SDP fields}
\maketitle

\section{Introduction}

{ The current estimates of the Star Formation Rate (SFR) of the Milky Way (MW) are uncertain mainly because we lack a general knowledge of the structure of our Galaxy. The most recent data suggest it is a two-armed, barred spiral with several secondary arms but the actual number and position of the arms is still unclear~\citep{Dame01,Dame11}. Moreover, the location of the Solar System in the Galactic Plane makes the definition of the overall Galactic structure, and consequently the heliocentric distance determination of star forming regions, even more difficult. Observations through the Galactic Plane are affected by source overlap along the line of sight and the optical and UV radiation emitted by young stars is absorbed or scattered by the Interstellar Medium (ISM) through the extinction process.}

{ Many SFR values have been measured during the years using different datasets and techniques, such as infrared (IR) photometry~\citep{Robitaille10}, which measures the light from young stars reemitted by the ISM in the IR, free-free emission~\citep{Murray10}, which measures the amount of photons required to produce the observed ionization of the HII regions, or high-mass star counts~\citep{Reed05}. }
{ The values range from 1~M$_\odot$/yr~\citep{Robitaille10} to 10~M$_\odot$/yr~\citep{Gusten82}. Recently~\cite{Chomiuk12} observed that, when normalizing all these measurements to the same Initial Mass Function (IMF), they converge to 1.9~M$_\odot/$yr.}

{ In recent years, a lot of progress has been made in identifying the position of many star forming regions~\citep{Russeil03,Benjamin05,Russeil11}, hence 
improving our knowledge of the structure of the Galaxy.
This has been made possible mostly by combining the information from line tracers with IR surveys, like GLIMPSE~\citep[``Galactic Legacy Infrared Mid-Plane Survey Extraordinaire'',][]{Benjamin03}, MIPSGAL~\citep[``A 24 and 70 Micron Survey of the Inner Galactic Disk with MIPS'',][]{Carey09}, and Hi-GAL~\citep[``Herschel Infrared Galactic Plane Survey'',][]{Molinari10a,Molinari10b}, which provide a wealth of data in the domain where the dust surrounding young stellar objects (YSOs) has the emission peak.}
{ In these bands it is therefore possible to study the quantity of young stars through the UV and optical light of the protostar which is absorbed, processed and reemitted by dust in the IR domain}. In particular, Hi-GAL mapped the Galactic Plane in $|\ell|<60^\circ$ and $|b|<1^\circ$ with the PACS~\citep{Poglitsch10} and SPIRE~\citep{Griffin10} instruments in parallel mode, in the 70, 160, 250, 350, and 500~$\mu$m bands. Given its spectral range and sky coverage, it provides a unique opportunity to study collapsing dust clouds and protostars, i.e. the early stages of star formation. Moreover, the survey is particularly sensitive to high mass stars (OB, $M>8 {\rm M_\odot}$) which regulate the ecology of our Galaxy as a whole.

In this paper, we estimate the SFR in the sky regions mapped during the Herschel Science Demonstration Phase (SDP), i.e. two $2^\circ\cdot2^\circ$ tiles centered on the Galactic Plane in $\ell=30^\circ$ and $\ell=59^\circ$. The $\ell=30^\circ$ field observes the Sagittarium and Perseus arms, while $\ell=59^\circ$ is centered on an inter-arm region~\citep{Russeil11}. Due to its location, the $\ell=30^\circ$ field is expected to be more active, in terms of star formation, than the $\ell=59^\circ$ field. A higher number of YSOs and HII regions are located in the $\ell=30^\circ$ area, including the W43 complex~\citep{Bally10}, a massive star forming region with an associated giant HII region. This gives us the opportunity to study the star formation processes both in a very active and in a quiescent environment. 

Due to the aforementioned difficulties in estimating the SFR of the MW, we do not have a reliable far IR star formation indicator. Many indicators have been calibrated on extragalactic star forming complexes and then applied to Galactic observations~\citep[see for example][]{Kennicutt98,Calzetti07,Kennicutt07}. 
{
Since one of the main goals of Galactic star formation studies is to derive a SFR estimator of the MW, hence test the validity of estimators calibrated on 
other galaxies,} we make use both of an extragalactic indicator~\citep{Li10} and of a star counting method to estimate the SFR in the two Hi-GAL SDP tiles, and then compare the results. This way, we both study the star formation activity in the two Hi-GAL fields and test the method in order to apply it to a larger sample when the whole Hi-GAL survey will be completed.
The paper is organized as follows: Sec.~\ref{sec:data} describes the dataset; Sec.~\ref{sec:sed} outlines the SED fitting, the sample selection criteria, and the estimate of the { candidate} YSOs physical parameters such as temperatures and masses; in Sec.~\ref{sec:sfr} we estimate the SFR with the extragalactic estimator and counting the young stars. The approximation adopted in this procedure and the quantifiable errors are discussed in Sec.~\ref{sec:errors}. Conclusions are summarized in Sec.~\ref{sec:conclusions}.


\section{The Dataset}~\label{sec:data}

The starting point of this study is the catalog of sources of the two Hi-GAL SDP fields~\citep{Elia10}. 
A new version of this catalog has recently been released to the Hi-GAL consortium, obtained with updated data cleaning procedures, mapmaking, calibration factors, extraction and photometry 
in the PACS and SPIRE bands. For further information about the Hi-GAL pipeline from raw data to map production we refer the reader to~\cite{Traficante11}. 
The algorithm used for the source detection is CuTEX (Curvature Thresholding EXtractor)~\citep{Molinari11} 
which double-differentiates the sky image and studies the variation of the curvature above a given threshold. { The identified source profiles are then fitted with a 2D elliptical Gaussian plus an underlying inclined planar plateau.}
This allows us to detect sources in the
presence of a variable background, as the one in the Galactic Plane, and to select not only point-like sources but compact objects. These two conditions
are crucial for us because we are interested in studying YSOs emitting in the PACS and SPIRE bands, which are still embedded. 
The considered sources have known kinematic distances~\citep{Russeil11} and are detected in at least three contiguous {Herschel} bands. This last condition is required to exclude spurious detections.
After these conditions have been applied, we have a total of 681 sources in $\ell=30^\circ$ and 316 in $\ell=59^\circ$. 
In order to better sample the SEDs and constrain the evolutionary stage, the 24~$\mu$m flux from MIPSGAL~\citep{Carey09} is also measured at the same position of the 70~$\mu$m band. 
For more information about the source extraction and distance determination we refer the reader to~\cite{Elia10} and~\cite{Russeil11}.

\section{SED fitting}~\label{sec:sed}

The physical parameters of the sources with at least three positive fluxes ($S_\lambda$) are estimated through a { modified black body} fit. The emission at wavelength $\lambda$ can be modeled as

\be\label{eq:greybody}
S_\lambda(\epsilon_0, T) = \epsilon_0 \left(\frac{\lambda}{\lambda_0}\right)^{-2}B_\lambda(T)
\ee
{ where $T$ is the source temperature, $\epsilon_0$ is the emissivity at the reference wavelength $\lambda_0=100\;\mu$m, the emissivity spectral index is set to 2 and $B_\lambda(T)$
is the blackbody at temperature $T$. The emissivity can be rewritten as 
\be
\epsilon_0 = \frac{M k_0}{d^2}
\ee

\noindent where M is the total mass of the source, $k_0$ is the mass opacity at wavelength $\lambda_0$ { per unit mass} and $d$ is the heliocentric distance.} 

We perform the fit using a Monte Carlo Markov Chain algorithm~\citep[MCMC][]{Lewis02} and estimate the temperatures and the total masses. We do not fit the SEDs with the grid of models from~\cite{Robitaille06}, as was done with the previous version of this catalog, because almost all the sources peak at wavelength $\lambda \geqslant 160~\mu$m which is the threshold identified by~\cite{Elia10} as the one {above} which the Robitaille models do not apply anymore. 
The fit is performed including only Herschel PACS and SPIRE bands. The associated error bars come mainly from calibration uncertainties and background removal. Since we are not analyzing point-like objects but embedded sources we make use of the calibration errors measured on extended emission which amount to 20$\%$ of the total flux for PACS and 15$\%$ of the flux for SPIRE. We also include a 10$\%$ statistical error coming from the background fluctuations. The two errors are added in quadrature and associated to the fluxes. 
In the present analysis we chose not to include the 24~$\mu$m flux in the fit since we are interested in the envelope emission. The 24~$\mu$m flux samples the internal source and this would require a two-component model, one for the envelope and one for the central star. 
We do not probe sources with $T<7$ K in our analysis because, in that temperature regime, the assumption of an optically thin regime implicit in the { modified black body} fitting breaks down. We would need a full radiative transfer model to account for optical depth effects and this goes beyond the scope of the present paper. 
The average temperature in the two fields are $<T_{l30}> = 17.5$ K and 
{ $<T_{\ell59}> = 14.3$ K}. The $\ell=59^\circ$ tile is colder on average, consistent with the fact that, being an inter-arm field, star formation is not very active. 


In order to estimate envelope masses, we put ourselves in the Rayleigh-Jeans (RJ) regime which is optically thin. 
Since the 500 $\mu$m band has a low resolution and is affected by source multiplicity issues, the 350 $\mu$m band is then the best compromise, so we estimate masses according to the formula   

\be\label{eq:mass}
M = \frac{S_{350}\; d^2}{k_{350}\; B_{350}(T)}
\ee

\noindent where $S_{350}$ is the flux at 350 $\mu$m, $d$ is the kinematic distance, $k_{350}= 0.07\; cm^2 g^{-1}$ is the mass opacity coefficient at 350 $\mu$m { per unit mass}~\citep{Preibisch93}. $T$ is the temperature estimated by means of Eq.~\ref{eq:greybody}. The measured masses range between 1 and $10^5$ M$_\odot$ in $\ell=30^\circ$ and between 0.1 and $10^4$ M$_\odot$ in $\ell=59^\circ$, with median values of 454 M$_\odot$ and 48 M$_\odot$ in $\ell=30^\circ$ and $\ell=59^\circ$, respectively. 

{
\subsection{Selection criteria}

Once we have fluxes, temperatures and masses of the whole sample of sources, we perform a further selection to ensure we collect only YSOs for the SFR estimate. First, we remove sources with a negligible emission in the PACS 70~$\mu$m band. Second, we identify and remove AGB stars using color-color criteria. 


The presence of a 70~$\mu$m counterpart has been demonstrated to be correlated with the internal luminosity of a protostar~\citep{Dunham08} and, moreover, we do not expect our survey to detect a 70~$\mu$m flux of diffuse ISM just heated by the interstellar radiation field. For those reasons, we consider as starless all sources without a detection at 70~$\mu$m.

One of the most challenging steps in cleaning our catalog is to make sure not to include AGB sources in our sample. Around $30\%-50\%$ of the mid-infrared sources are estimated to be AGBs by~\cite{Robitaille08} using the GLIMPSE survey.
AGB sources also emit in the IR and can be disentangled from YSOs by applying color-color criteria. We make use of the criteria developed from~\cite{Martinavarro12} in the Herschel bands. According to these authors, AGBs are located in the color-color diagram where the following is verified:

\be
0.2 <  \displaystyle{\frac{\log( \frac{S_{70}}{S_{160}})}{\log( \frac{S_{70}}{S_{250}})}}  <  0.9 ,\; \; \; \; \;  1.2  < \displaystyle\frac{\log( \frac{S_{70}}{S_{350}})}{\log( \frac{S_{160}}{S_{350}})}  <  3.3
\ee

The percentage of AGBs among sources detected at 70~$\mu$m is 47$\%$ in $\ell=30^\circ$ and 52$\%$ in $\ell=59^\circ$, which is consistent with what was found in the GLIMPSE survey. 
All the sources satisfying these color criteria are removed from the sample. 

In order to identify starless objects which are gravitationally bound, i.e. prestellar sources, we apply the method used by~\cite{Giannini12} on the Vela-C molecular cloud. Instead of the virial mass, they make use of the Bonnor-Ebert mass, ${ M_{BE} \sim 2.4R_{BE}a^2/G}$, where $ {R_{BE}}$ is the Bonnor-Ebert radius which can be approximated by the actual radius of the source, $R_{BE} = \delta/2$ where $\delta$ is the diameter. $G$ is the gravitational constant and ${ a=\sqrt{k_BT/\mu}}$. Here, $k_B$ is the Boltzmann constant, $T$ is the source temperature and $\mu$ is the mean molecular weight. 
The source diameters ($\delta$) are estimated from the 250~$\mic$ map by approximating the ellipse provided by CuTEX as a circle with the same area and deconvolving it by the instrumental beam. The choice of the 250 $\mu$m has been done because all sources are detected in this band. The size distribution
is shown in Fig.~\ref{fig:size_plots}. In both tiles the majority of sources have
$\delta > 0.1$~pc meaning that they are essentially clumps (see e.g.~\cite{Kauffmann10} and references therein), as is also confirmed by their mass distribution.

\begin{figure}[!t]
\begin{center}
\includegraphics[width=0.24\textwidth]{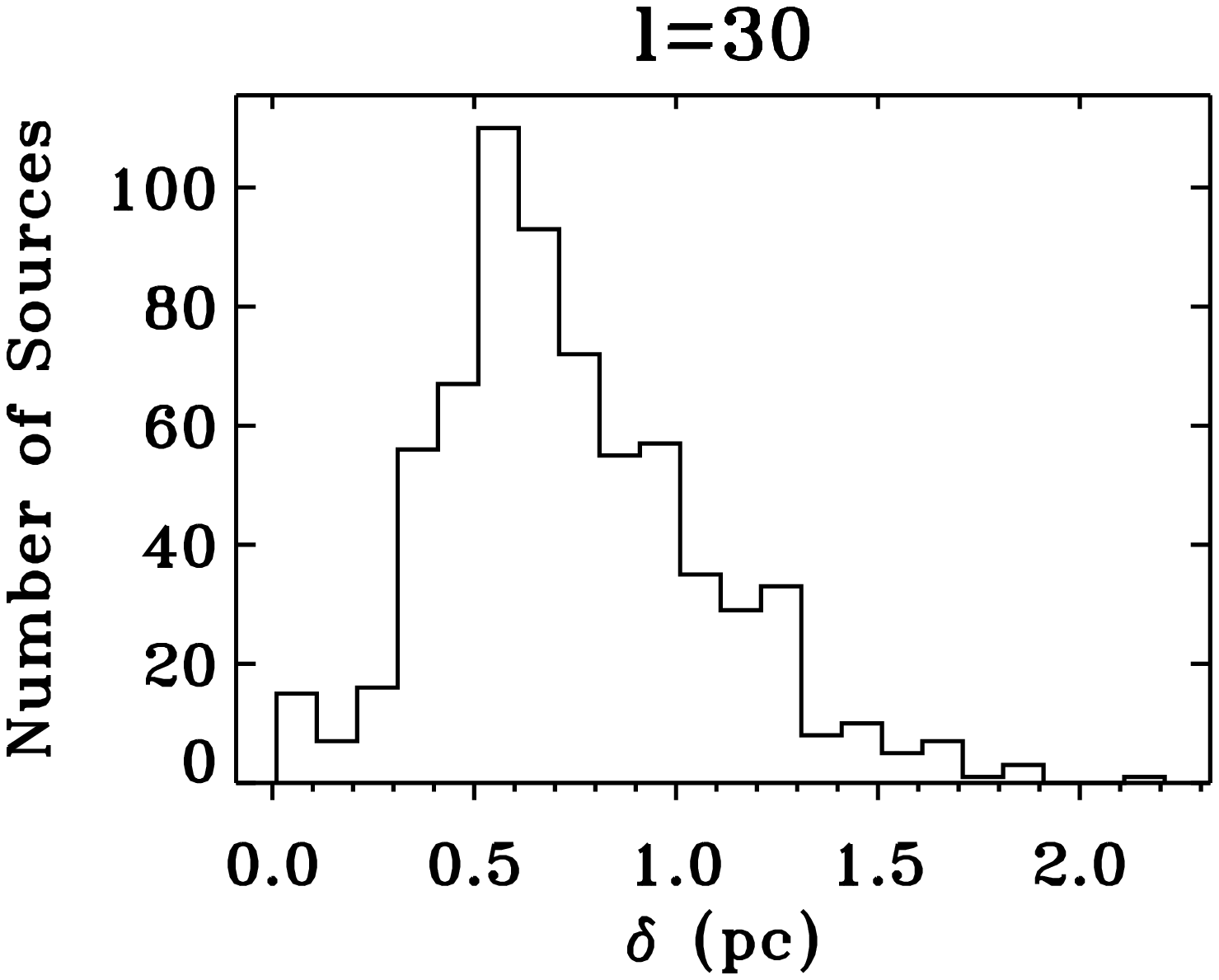}
\includegraphics[width=0.24\textwidth]{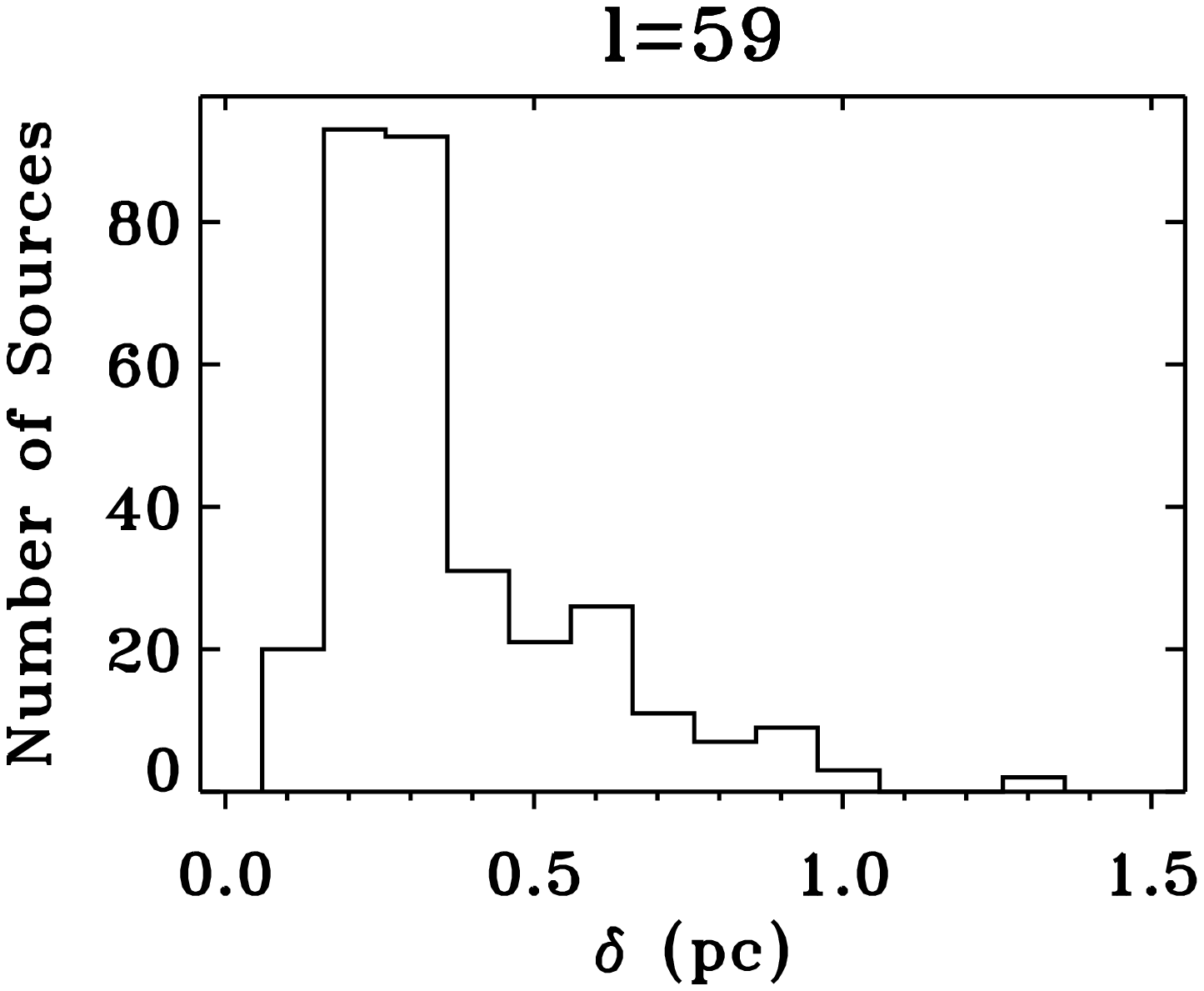}
\caption{Diameter distribution of all sources detected in three adjacent Herschel bands, estimated from the 250 $\mu$m map. }
\vspace{0.5cm}
\label{fig:size_plots}
\end{center}
\end{figure}

Following the criteria in~\cite{Giannini12} we identify sources with $M/M_{BE}\geqslant 0.5$ as gravitationally bound.    

The temperature distribution of the prestellar and protostellar sources is reported in Fig.~\ref{fig:temp_beta_plots}. This figure clearly shows two populations of objects, a cold one (red histogram), mostly prestellar, which peaks around 11 K and a warm one (blue histogram), mostly protostellar, which peaks around 20 K. The protostellar sources constitute the final sample used in the following analysis. 

\begin{figure}[!t]
\begin{center}
\includegraphics[width=0.5\textwidth]{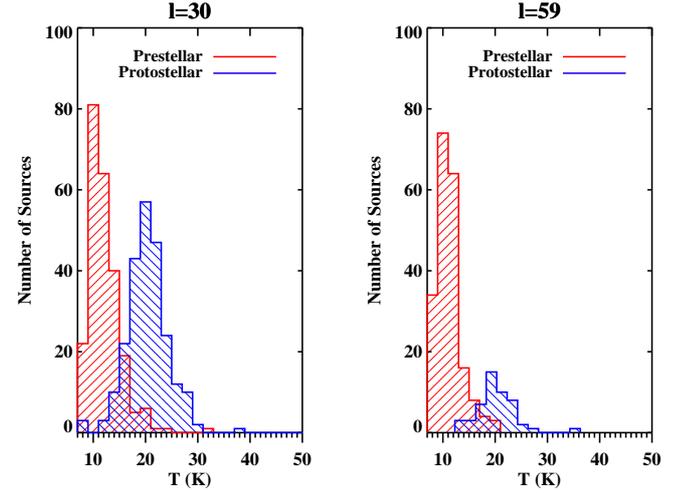}
\caption{Temperature distribution of the considered YSO { candidates} and prestellar cores estimated by means of a { modified black body} fit from 70~$\mu$m to 500~$\mu$m. }
\vspace{0.5cm}
\label{fig:temp_beta_plots}
\end{center}
\end{figure}

{ Two recent papers~\citep{Battersby11,Paradis10} have measured the temperature of these fields on a pixel-by-pixel basis, smoothing the signal of all wavelengths to 35.6'', i.e. the resolution of the Herschel 500~$\mu$m band. Their results don't show the same bimodal distribution as ours because, in our case, we keep the original resolution of each band and, therefore, we identify and collect compacts objects and we are sensitive to the temperature variations on these smaller scales. }

After applying those criteria, the number of identified { candidate} YSOs is 235 in $\ell=30^\circ$ and 50 in $\ell=59^\circ$. { Table~\ref{tab:data} gives the coordinates, fluxes and distances of the 285 selected sources. }

\begin{table*}[!t]
\begin{center}
\space
\caption{List of { candidate} YSOs in the SDP fields, their MIPS and Hi-GAL photometry and distances}
\label{tab:data}
\begin{tabular}{c | c  c  | c  c c c c c | c }
\hline
Source Name & Glon & Glat & F(24$\mu$m)  & F(70$\mu$m)  &  F(160$\mu$m)  & F(250$\mu$m)  & F(350$\mu$m)  & F(500$\mu$m)  & d\\
  & deg & deg & Jy & Jy & Jy & Jy & Jy & Jy & kpc\\
\hline
\hline
 G029.0540+0.5062 &   29.0540 &    0.5062 &       0.03  &  1.78 &  11.68 &  11.82  &  6.73  &  4.27  &  8.90 \\
 G029.4191+0.6303 &   29.4191 &    0.6303  &    0.31  &  7.36  &  7.89  &  4.56  &  2.96  &  1.96  & 13.50 \\
 G029.1367+0.4649 &   29.1367 &    0.4649  &    5.01  &  7.77  & 14.59  &  9.78  &  7.41  &  2.45  & 13.10 \\
 G029.1612+0.4743 &   29.1612 &    0.4743    &    2.55  &    4.33  &   18.46   &  19.06  &   13.34   &   6.51 &    13.10  \\
 G029.1581+0.4637 &   29.1581 &    0.4637 &       0.24  &   25.74   &  32.29   &  18.80   &  13.96    & -   &         13.10 \\
 G029.1383+0.4195 &   29.1383 &    0.4195  &      0.44   &   5.80    &  8.15    &  3.53    &  1.63   &   0.97  &   14.20 \\
 G029.1538+0.4161 &   29.1538 &    0.4161   &      0.32  &   11.27  &   12.53   &   4.09   &   4.07      &  -   &     13.30 \\
 G029.8533+0.5491 &   29.8533 &    0.5491   &     0.08   &   7.64   &  11.83   &   6.44   &   3.91    &  2.64    &  4.50 \\
 G029.0238+0.0842 &   29.0238 &    0.0842   &     8.14   &   2.30   &  18.14   &  18.78  &   10.33    &  8.40     & 9.90 \\
 G029.8853+0.4918 &   29.8853 &    0.4918   &     0.05   &   1.72    &  5.10    &  5.71   &   5.00   &   2.25 &    13.40 \\
 G030.4278+0.7694  &  30.4278 &    0.7694    &    0.17   &   4.20    &  5.54    &  2.86    &  1.36   &   0.50    &  11.50 \\
 G029.1172+0.0889 &   29.1172 &    0.0889    &   44.05   &  69.27   &  72.11  &   41.44   &  30.04   &  16.74   &   9.30 \\
\hline
\end{tabular}
\end{center}
\footnotesize{Only a portion of this table is shown here to demonstrate its form and content. A machine-readable version of the full table is available at this \href{http://cdsarc.u-strasbg.fr/viz-bin/Cat?cat=J%2FA%2BA&categ=J%2FA%2BA&}{link}.}
\end{table*}
\space

{ The distance distribution of the selected sources is shown in Fig.~\ref{fig:dist}. In $\ell = 30^\circ$ (left panel), most of the sources cluster, as expected, in the Scutum spiral arm which crosses the line of sight twice (at d$~\sim5.5$ kpc (near) and d$~\sim9$ kpc (far)) and in the Sagittarius arm (at d$~\sim3.5$ kpc (near) and d$~\sim11$ kpc (far)). The peak centered in d$~\sim13$ kpc corresponds to sources located in the Perseus arm. In $\ell = 59^\circ$ (right panel) the line of sight is tangent to the Sagittarius arm (d$~\sim3-7$ kpc) where most of the sources are located. 
For further information about the Galactic distribution of Hi-GAL sources in the SDP fields and for distance derivation we refer the reader to~\cite{Russeil11}.}

\begin{figure}[th]
\begin{center}
\includegraphics[width=0.5\textwidth,angle=0]{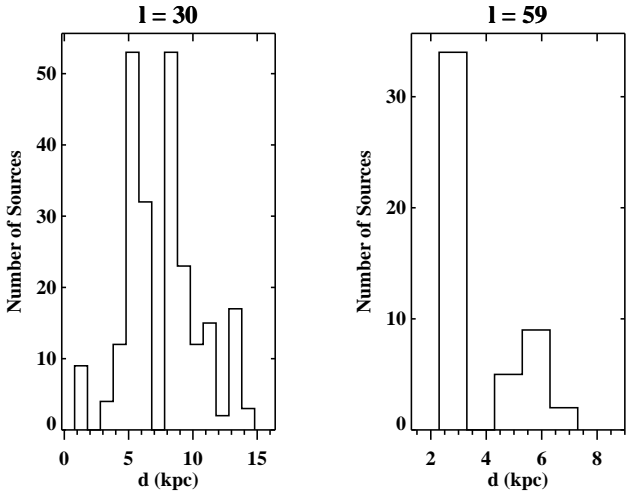}
\caption{{ Distance distribution of the selected { candidate} YSOs in SDP fields.}}
\label{fig:dist}
\end{center}
\end{figure}

The color-color plots of the final sample are shown in Fig.~\ref{fig:cc_plots}. { Each axis of these figures report the magnitude difference between two bands according to the formula $[\lambda_1 - \lambda_0] = m_1 - m_0 = -2.5\log(\frac{S(\lambda_1)}{S(\lambda_0)})$ where $S(\lambda)$ is the flux at the band $\lambda$ and $m$  is its magnitude.}
The left panel shows a subsample of the right panel because some of the protostellar sources are detected at 70~$\mu$m but not at 24~$\mu$m.

\begin{figure}[th]
\begin{center}
\includegraphics[width=0.5\textwidth]{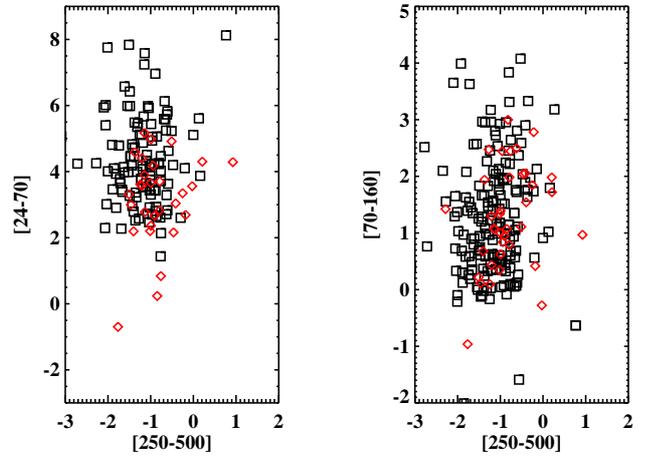}
\caption{{ Color-color diagrams for all the protostellar sources in the $\ell=30^\circ$ (black squares) and $\ell=59^\circ$ (red diamonds) fields. The color distribution is consistent with embedded sources which can be fitted only with a { modified black body} and not with an embedded ZAMS model~\citep{Molinari08}.}}
\label{fig:cc_plots}
\end{center}
\end{figure}


}

\section{SFR estimate}\label{sec:sfr}

The SFR in the SDP fields is estimated using two independent techniques, one based on an IR extragalactic estimator and the other on source counting. Results for the tiles as a whole are reported in Tab.~\ref{tab:result}. 

\begin{table}[t]
\begin{center}
\space
\caption{SFR estimates}
\label{tab:result}
\begin{tabular}{l c | c  c  }
\hline
Method & Field & SFR$_{tile}$  &  SFR$_{MW}$   \\
 & & (M$_\odot$/yr) & (M$_\odot$/yr)  \\
\hline
\hline
\multirow{2}{*}{Star counts}& $\ell = 30$ & $ {(9.5\pm4.3)\cdot10^{-4}}$ & -- \\
& $\ell = 59$ & $  {(1.6\pm0.7)\cdot10^{-4}}$& -- \\
\hline
\multirow{2}{*}{IR estimator}& $\ell = 30$ & ${(2.4 \pm 0.4)\cdot 10^{-4}}$ &  ${0.71\pm0.13}$ \\
&$\ell = 59$ & ${(2.6 \pm 1.1) \cdot 10^{-6}}$ & ${0.10\pm0.04}$ \\
\hline
\end{tabular}
\end{center}
\footnotesize{}
\end{table}
\space

\subsection{Source counts}

\begin{figure}[!t]
\begin{center}
\includegraphics[width=0.24\textwidth]{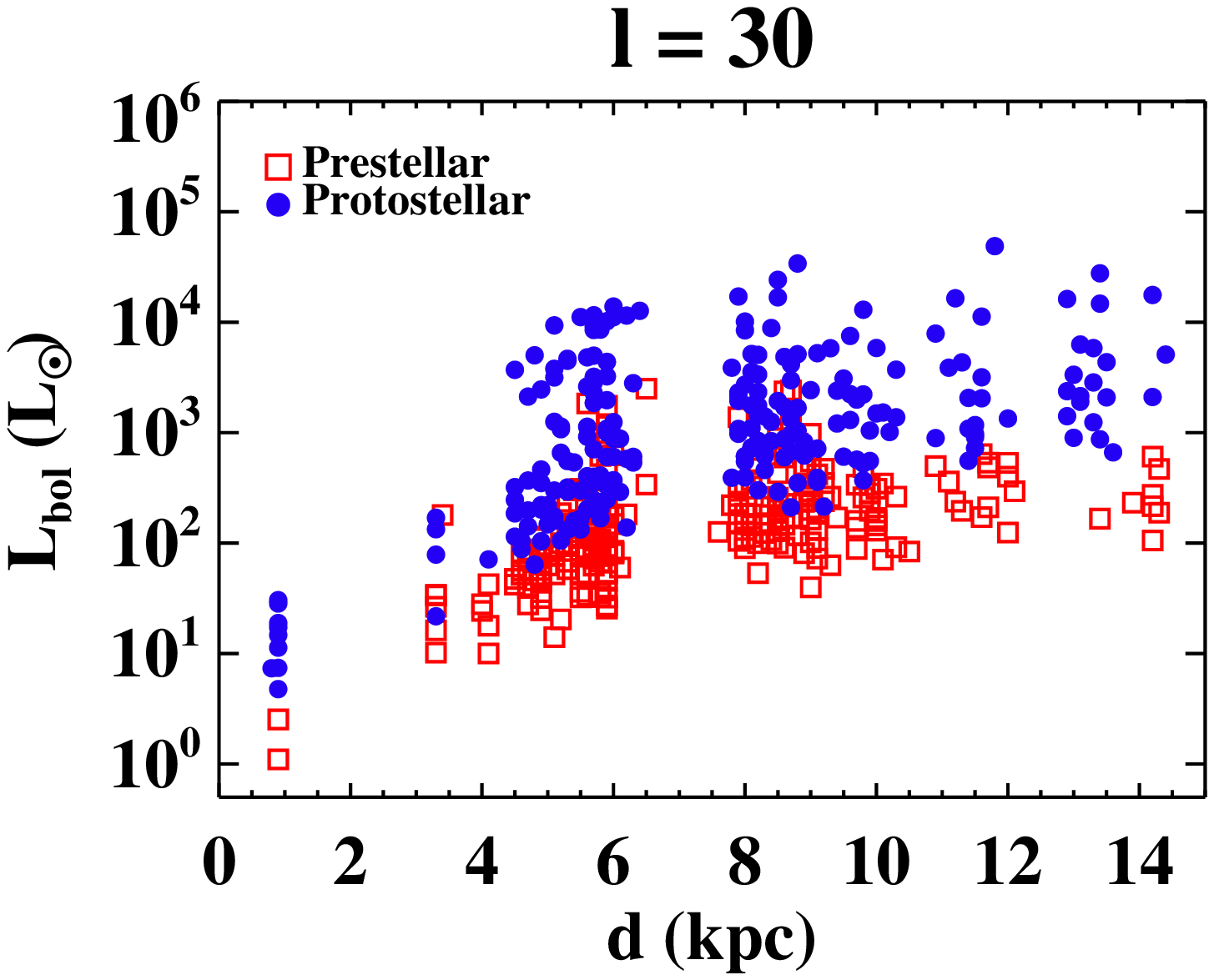}
\includegraphics[width=0.24\textwidth]{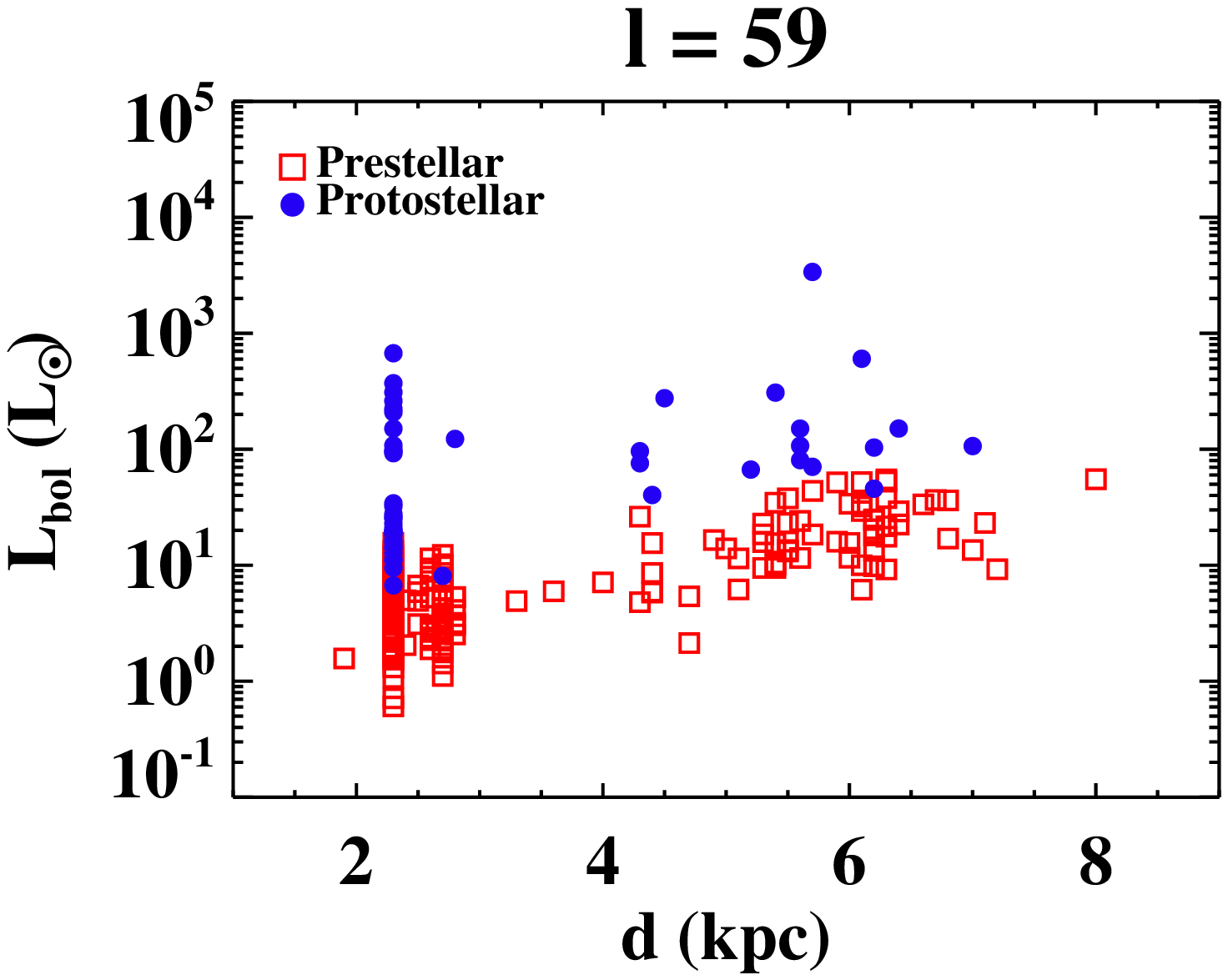}
\caption{ Galactic distribution of the bolometric luminosity of prestellar (red squares) and protostellar (blue circles) sources in the SDP fields. }
\label{fig:dist_lbol}
\end{center}
\end{figure}

Starting from fluxes and distances we estimate the bolometric luminosities ($L_{bol}$) of our sources according to the formula

\be\label{eq:lbol}
L_{bol} = 4\pi d^2\int_{\lambda_{min}}^{\lambda_{max}} S_\lambda d\lambda
\ee

\noindent where $\lambda_{min} = 24\;\mu$m, $\lambda_{max} = 500\;\mu$m, $d$ is the heliocentric distance and $S_\lambda$ is the source flux at the wavelength $\lambda$.
{ The prestellar and protostellar source distributions of the $L_{bol}$ in the Galaxy towards the two considered tiles is shown in Fig.~\ref{fig:dist_lbol}. }

We then build the $M_{env}-L_{bol}$ diagram where $M_{env}$ is the envelope mass. This kind of diagrams provides information on the evolutionary stage of the source and a prescription of when it will join the Zero Age Main Sequence (ZAMS hereafter) and with which final mass~\citep{Molinari08}. The source evolves along defined tracks according to the chosen evolutionary model. We adopt the model of collapse in turbulence-supported dust cores~\citep{McKee03}. 
{ According to this model, the evolution of high-mass YSOs has two main phases. In a first phase, as soon as the initial cloud starts collapsing, the luminosity of the embedded star increases in an accelerated fashion and the initial mass envelope decreases because part of the material feeds the central core and part is expelled through molecular outflows. The path followed by the protostar in the $M_{env}-L_{bol}$ diagram is almost vertical. At the end of this phase, the star reaches the ZAMS, or is very close to it, and is surrounded by an HII region. The average time of this phase is 2.7${\cdot10^5}$ yr for an initial envelope mass of 13.5~M$_\odot$~\citep{Molinari08}.
In the second phase, after the end of the accelerating accretion, the remaining envelope mass is partially drained by other objects forming in the same clump and part of it is expelled through molecular outflows. An accretion disk is also present, so part of the initial cloud still keeps feeding the central star. The envelope clean-up phase ends when the object is visible in the optical band and the initial cloud has completely disappeared. Since the luminosity in this phase remains almost constant, this results in an horizontal track in the diagram. The total average evolutionary time for the entire path is 3.5${\cdot10^6}$ yr for an initial envelope mass of 13.5~M$_\odot$. For more detailed information about the evolutionary model we refer the reader to~\cite{McKee03} and~\cite{Molinari08}.}

\begin{figure*}[!t]
\begin{center}
\includegraphics[width=0.45\textwidth]{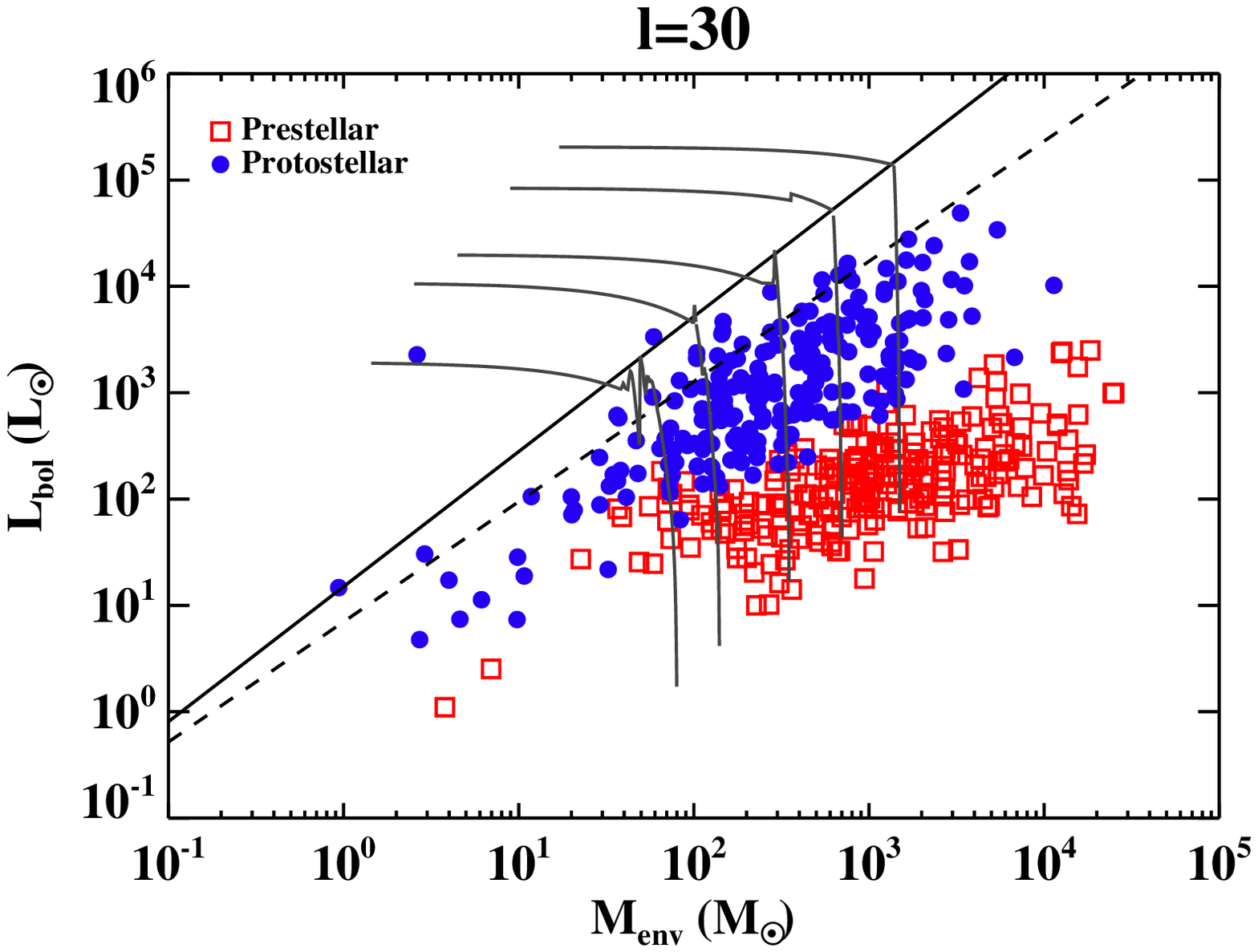}
\includegraphics[width=0.45\textwidth]{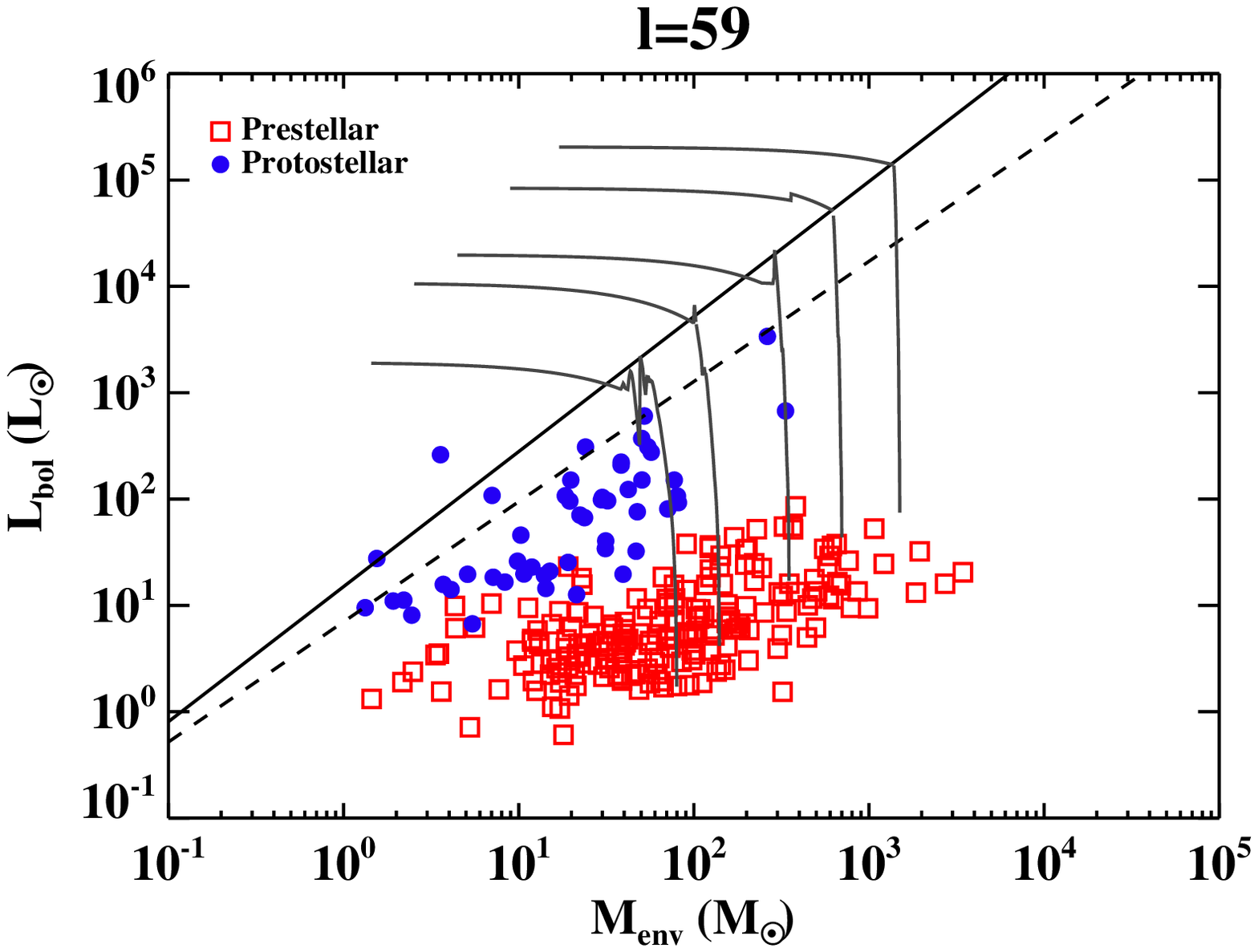}
\caption{{ $M_{env}-L_{bol}$ diagram of the sources in the $\ell=30^\circ$ (left panel) and $\ell=59^\circ$ (right panel) fields for the prestellar and protostellar cores shown in Fig.~\ref{fig:temp_beta_plots}. The solid black line and the dashed black line are the best log-log fit of the high-mass counterparts of the Class I and Class 0 sources in the low-mass regime, as found in~\cite{Molinari08}, respectively. }}
\label{fig:lsol_msol}
\end{center}
\end{figure*}

Since the sources we are considering are still deeply embedded, we can assume that $M$ corresponds to the $M_{env}$ derived in Sec.~\ref{sec:sed}. The $M_{env}-L_{bol}$  diagrams for the two tiles are shown in Fig.~\ref{fig:lsol_msol}.
{ The solid black line is the best log-log fit of the high-mass counterpart of the low-mass Class I regime~\citep{Molinari08}, while the dashed black line is the high-mass counterpart of the best log-log fit of the low-mass Class 0 regime. In the high-mass domain the subdivision of young stars in classes is a conventional extrapolation from low-mass stars. 
{ Since high-mass clumps contain unresolved structures which might fragment into more than one high-mass star, and continue to accrete even after entering the main sequence, one has to be careful about this subdivision.}} 
The majority of the sources in our sample are below this line, meaning that these objects are still in an early evolutionary phase as already argued from the color-color plots in Fig.~\ref{fig:cc_plots}. The average uncertainties on luminosities and masses in $\ell=30^\circ$ are $<\sigma_L/L> = 0.2$ and $<\sigma_M/M> = 0.4$, while in  in $\ell=59^\circ$ are $<\sigma_L/L> = 0.4$ and $<\sigma_M/M> = 0.7$. These values have been estimated through the error propagation of fluxes and distances.

In order to estimate the SFR from the evolutionary tracks provided by~\cite{McKee03}, we bin these diagrams into two-dimensional histograms. Each bin has an associated formation time which depends on the source stage and mass. 
{ Since high-mass objects keep accreting even after they join the ZAMS, the formation timescales used in our analysis cover the entire evolutionary path, from the beginning of the { accelerating accretion phase to the end of the envelope clean-up phase, i.e. when the envelope has disappeared and the object is visible in the optical bands.} They are ${t_f  = (2.1,\;2.7,\;3.5,\;4.8,\;4.5)\cdot10^6}$~yr for sources with initial envelope masses in the bins (80, 140, 350, 700, 2000) M$_\odot$, respectively~\citep{Molinari08}}. Due to Hi-GAL sensitivity limits, we are not confident in the detection of objects with M$_{env}<{\rm 10\; M}_\odot$. These sources are then excluded from further analysis and the missing contribution to the SFR, coming from the low-mass regime, is estimated in Sec.~\ref{sec:errors}.

The SFR from { candidate} high-mass YSOs in each tile is obtained by summing up the final masses and dividing them by the associated formation time:

\be
SFR = \sum_{i = 1}^{N_{Class}} \sum_{j=1}^{N_{Mass}} n_{M}(i,j) M_{ZAMS}(j) / t_f(i) \;\;\;{\rm M_\odot/yr}
\ee

\noindent where ${ N_{Class} = 3}$ { (counterparts of low-mass Class 0, I, II)}, $N_{Mass} = 5$ are the number of classes and of initial masses, respectively. $n_{M}(i,j)$ is the number of sources of Class i with envelope mass j and $M_{ZAMS}(j)$ is the expected final mass of sources of initial envelope mass j when they reach the ZAMS.  
{ The SFR obtained with this method is ${ \rm{(9.5\pm4.3)\cdot10^{-4}\;M_\odot/yr}}$ in $\ell=30^\circ$ and ${ {\rm (1.6\pm0.7)\cdot10^{-4}\;M_\odot/yr}}$ 
in $\ell=59^\circ$.}
We do not extrapolate these estimates to the whole MW because they have been calculated through a counting procedure, so they are very local values. { Since YSOs generally join the ZAMS when the { accelerating} accretion phase has been completed, we report also the SFR estimated at the end of this phase, so with shorter evolutionary timescales, in order to provide an upper limit to the estimates. The average evolutionary timescales of the main accretion phase for the mass bins mentioned before are ${t_f  = (4.5,\;3.7,\;2.7,\;2.1,\;1.5)\cdot10^5}$~yr and the correspondent SFR would be ${\rm (1.6\pm0.7)\cdot10^{-2}\;M_\odot/yr}$ in $\ell=30^\circ$ and ${\rm (7.9\pm3.6)\cdot10^{-4}\;M_\odot/yr}$ in $\ell=59^\circ$.}

\subsection{ Monochromatic estimator at 70~$\mu$m }

Another way to study the SFR of a sample of sources with known infrared luminosities is to make use of an extragalactic estimator. 
{ Dust surrounding the forming stars absorbs UV radiation and re-radiates it in the IR.}
Therefore, optical/UV based indicators are not reliable in the MW because of the high extinction generated by interstellar dust along the line of sight. On the contrary, IR emission is a reliable tracer of the photons produced by the forming source.
\cite{Lawton10} finds that, among all the IR bands, the 70~$\mu$m band is the most reliable 
indicator of star formation because it is able to account for IR emission from HII regions~\citep[see for example][]{Tibbs12,Faimali12}.

We then consider the 70~$\mu$m monochromatic estimator developed by~\cite{Li10}. They measure the 70~$\mu$m luminosities of extragalactic star forming regions and calibrate them on their previous SFR estimator based on 24~$\mu$m and H($\alpha$) luminosities~\citep{Calzetti07}. The conversion factor between total luminosity at 70~$\mu$m and the SFR is $1.067\cdot10^{43}$ erg/s. { The SFR obtained with this method is ${ \rm{(2.4\pm0.4)\cdot10^{-4}\;M_\odot/yr}}$ in $\ell=30^\circ$ and ${ {\rm (2.6\pm1.1)\cdot10^{-6}\;M_\odot/yr}}$  in $\ell=59^\circ$.}

Since the estimator is built averaging extragalactic star forming complexes, we extrapolate these values to the whole Galaxy modeling it as a disk in which sources are uniformly distributed within the volume. The observed fields are then slices of this volume, with the vertex centered on the Sun and an aperture angle of 2$^\circ$, i.e. the dimension of the tiles. The SFR value of the single tiles is then extrapolated to the whole MW through the equation:

\be
SFR_{MW} = SFR_{t} \frac{V_{MW}}{V_{t}}
\ee

\noindent where

\bea\label{eq:vol_ratio}
&& V_{MW}   = \pi R^2 h_{MW}\\
&& V_{t}  = \frac{d_t^2 \alpha}{2} z_{t}\nonumber
\eea
 
\begin{figure}[!t]
\begin{center}
\includegraphics[width=0.45\textwidth]{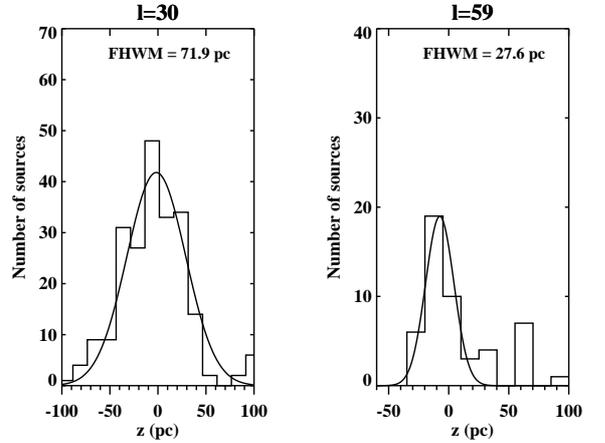}
\caption{z distribution of the Hi-GAL protostars in the $\ell=30^\circ$ and $\ell=59^\circ$ fields. The z value adopted to estimate the tile volumes is the FWHM of the Gaussian fit. { The second peak at $z\sim60$~pc in $\ell=59^\circ$ is due to a concentration of YSOs in the Vulpecula OB association~\citep{Billot10}}. }
\label{fig:z_plots}
\end{center}
\end{figure}

{  In the previous equations, $V_t$ and $V_{MW}$ are the volumes of the tile and of the whole MW, respectively. $h_{MW}= 0.3$~kpc is the thickness of the Galactic disk and $R=15$~kpc is its radius, $d_t$ is the average heliocentric distance of the { candidate} YSOs in the tile (7.6 kpc in $\ell=30^\circ$ and 3.3 kpc in $\ell=59^\circ$), $\alpha=2^\circ$ is the aperture angle and $z_{t}$ is the distance of the sources from the Galactic plane.} Following~\cite{Paladini04}, we estimate $z_{t}$ by means of a Gaussian fit of the $z = d \sin(b)$ distribution of the sources, where $b$ is their Galactic latitude. The FWHM of the fitted Gaussian curve is the $z_{t}$ value we assume for all the protostars in the same tile. 
{ The fits are shown in Fig.~\ref{fig:z_plots}. We obtain $z_{\ell30} = 71.9$~pc and $z_{\ell59}=27.6$~pc. These values are in agreement with the distribution of 456 Galactic HII regions reported in~\cite{Paladini04}. }

{ We obtain a SFR${ _{MW} = 0.71\pm0.13\,{\rm M_\odot/yr}}$ in $\ell=30^\circ$ and SFR${_{MW} =0.10\pm0.04\,{\rm M_\odot/yr}}$ in $\ell=59^\circ$}. As expected, the extrapolation from $\ell = 30^\circ$ gives more reasonable results since it is an area with active star forming regions and, therefore, it has characteristics similar to the star formation dominated galaxies used to calibrate the estimator. In fact, the value of SFR$_{MW}$ from $\ell=30^\circ$ is in good agreement with previous results in the MW obtained using different estimators and datasets~\citep{Chomiuk12} {and it falls in the range measured by~\cite{Robitaille10} using the {\em Spitzer}/IRAC GLIMPSE survey ($0.68-1.45\; M_\odot/$yr).}
{ In the previous extrapolation of the local SFR to the whole Galaxy we set the overall tile distance $d_t$ (Eq.~\ref{eq:vol_ratio}) to the average distance of the sources in the considered tile.  
When we set $d_t$ to the distance of the further source in the tile ($d_t$ = 14.4 kpc in $\ell=30^\circ$ and $d_t$ = 7 kpc in $\ell=59^\circ$), i.e. the distance in which we assume our sample to be complete, we obtain SFR$_{MW} = 0.20\pm0.04\,{\rm M_\odot/yr}$ in $\ell=30^\circ$ and SFR$_{MW} =0.02\pm0.01\,{\rm M_\odot/yr}$ in $\ell=59^\circ$.}

\section{Catalog completeness}\label{sec:errors}

\begin{figure}[!t]
\begin{center}
\includegraphics[width=0.5\textwidth]{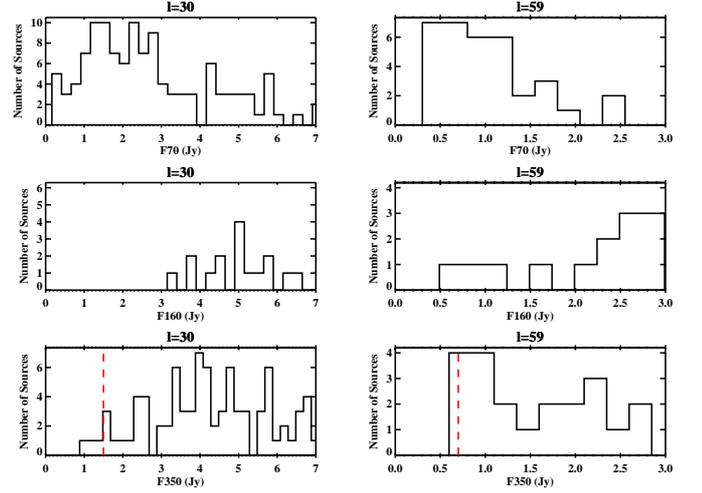}
\caption{Distribution of the { candidate }YSOs 70~$\mu$m, 160~$\mu$m and 350~$\mu$m fluxes in the SDP fields. These bands are particularly important because the 70~$\mu$m fluxes are used as indicator of star formation activity, they 160 $\mu$m is located in the peak of the SED and the 350 $\mu$m fluxes determine the masses through Eq.~\ref{eq:mass}. The vertical red-dashed lines in the bottom panels indicate the sensitivity limit used to calculate the completeness of our sample. They correspond to { 1.5 and 0.7 Jy} in $\ell=30^\circ$ and $\ell=59^\circ$, respectively. 
 }
\label{fig:completness_plots}
\end{center}
\end{figure}

The catalog completeness provides us with the information about how many sources we can miss with fluxes fainter than a given threshold. This translates into how much mass and luminosity we are missing when calculating the SFR. Since the masses have been estimated from the 350 $\mu$m flux (Eq.~\ref{eq:mass}), we study the distribution of the fluxes in this band to identify the lower detection threshold. In Fig.~\ref{fig:completness_plots} we show a zoom of the 70~$\mu$m (top line), 160 $\mu$m (central line) and 350 $\mu$m flux distribution (bottom line) of the { candidate} YSOs in the two tiles. We can see that below a given flux {value} less and less sources are detected and that this threshold is lower in $\ell=59^\circ$ than in $\ell=30^\circ$. The threshold value depends mostly on the instrumental sensitivity, on the detection algorithm and on the observed field. The more the field is populated, as in the $\ell=30^\circ$ case, with bright sources and variable background, the more difficult it is for the detection algorithm to detect faint objects. We chose as a flux sensitivity threshold at 350 $\mu$m the values right before the first peak of sources. These limit fluxes are { 1.5~Jy and 0.7~Jy} for $\ell=30^\circ$ and $\ell=59^\circ$, respectively. These values are very similar to the ones estimated by~\cite{Molinari10a} through synthetic source experiments in the old version of the catalog.   
In order to translate those values into the amount of missing mass, we estimate the mass corresponding to those limit fluxes using Eq.~\ref{eq:mass} where we assign, as distance, the median distance detected in the field and, as temperature, the average temperature of protostars in that field. The conservative choice of the median distance associated to the limit flux allows us to include even the faintest objects in the estimate of missing mass. 
{ The envelope masses corresponding to the limit flux then turn out to be ${M_{env}^{lim} = 66\;{\rm M_\odot}}$ and ${M_{env}^{lim} = 6.5\;{\rm M_\odot}}$ in $\ell = 30^\circ$ and $\ell = 59^\circ$, respectively. According to~\cite{Molinari08} and~\cite{Saraceno96}, this translates into a final stellar mass of  ${M_{\ell30}^{lim} = 6.5\;{\rm M_\odot}}$ and ${M_{\ell59}^{lim} = 0.8\;{\rm M_\odot}}$.
The number of sources expected with $M<M_{lim}$ is estimated with the~\cite{Kroupa01} Initial Mass Function (IMF). 
The IMF is normalized to the counts in the { bin of lowest envelope masses ($80\;M_\odot<\Delta M_0< 140\;M_\odot$) where our detection is complete}, according to the formula 
\be
N(M_{min}) \simeq N(M_0)\frac{\Delta M_{min}}{\Delta M_0}\left(\frac{M_{min}}{M_0}\right)^{-\gamma}
\ee
\noindent where $N(M_{min})$ is the number of stars in the bin $\Delta M_{min} = [0.1,0.5]\;M_\odot$,  $N(M_0)$ is the number of stars in the bin $\Delta M_0$ and $M_0$ is the limit mass of the final star $M^{lim}$ in each tile. $\gamma = 2.3$ for $M>0.5\;{\rm M_\odot}$ and $\gamma = 1.3$ for $0.1\;{\rm M_\odot}<M<0.5\;{\rm M_\odot}$. We then calculate the number of missing stars until $M_{min}=0.1\;{\rm M_\odot}$, which we assume to be the minimum value for a YSO mass. 
By means of this procedure we estimate that, considering 0.1 ${\rm M_\odot}$ as lower limit for a final star mass, we are not detecting $\sim55\%$ of the total number of sources in $\ell=30^\circ$ and $\sim3\%$ in $\ell=59^\circ$. 
The average timescale for a low-mass YSO population is $\tau \sim 2$ Myr (see for example~\cite{Evans09,Covey10,Lada10}). By using again the star counts method, we can then calculate the amount of missing SFR as SFR$_{M < M(lim)} = M_{min} \cdot N(M_{min}) / \tau$. 
We obtain SFR$_{M < M(lim)} = 1.0\cdot10^{-5}\;{\rm M_\odot/yr}$ and SFR$_{M < M(lim)} =3.9\cdot10^{-8}\;{\rm M_\odot/yr}$ in $\ell=30^\circ$ and $\ell=59^\circ$, respectively. Those values are negligible with respect to the estimates of the SFR obtained with both methods. }

\section{Summary and conclusions}\label{sec:conclusions}

We estimate the SFR in the two best studied Hi-GAL fields making use of the most updated version of the source catalog. Due to the spectral range, sensitivity and sky coverage of our dataset and because of the extraction algorithm, we are particularly sensitive to the very early stages of high-mass star formation. Moreover, the two Hi-GAL fields in this study cover both Galactic arms and inter-arm regions and this gives us the chance to study the SFR both in a very active and in a more quiescent field. { Thanks to color-color criteria and to the knowledge of heliocentric distances, we are able to remove AGBs stars, keeping only the protostellar sources. }

In this work we make use both of an extragalactic monochromatic estimator at 70~$\mu$m and of a source counting procedure and compare the results. { The SFR estimates for the two tiles are ${(9.5\pm4.3)\cdot10^{-4}\;{\rm M_\odot/yr}}$ and ${(1.6\pm0.7)\cdot10^{-5}\;{\rm M_\odot/yr}}$ with the star counts method in $\ell=30^\circ$ and $\ell=59^\circ$, respectively. Results with the IR estimator are ${(2.4\pm0.4)\cdot10^{-4}\;{\rm M_\odot/yr}}$ and ${(2.6\pm1.1)\cdot10^{-6}\;{\rm M_\odot/yr}}$ in $\ell=30^\circ$ and $\ell=59^\circ$, respectively. Those values are in good agreement in $\ell=30^\circ$, which is an active field. }

We also estimated that our catalog is complete { at $\sim45\%$ in $\ell=30^\circ$ and $\sim97\%$ in $\ell=59^\circ$}. Because of the instrument characteristics, in $\ell = 30^\circ$ we miss a large population of low-mass stars while the $\ell = 59^\circ$ field is closer and less populated, so the sampling is more complete.  
{ The contribution to the Galactic star forming activity coming from non detected low-mass stars is estimated to be at least more than one order of magnitude lower than the SFR from high-mass objects obtained using both methods.} 

The extragalactic indicator is, by construction, an average value of extragalactic star forming regions, so it provides SFR estimates which are meant to be calculated on whole galaxies. Therefore, it makes sense to extrapolate the local results of the tiles to the whole MW. When applying this method to the $\ell=30^\circ$ tile, which has a similar star formation activity as the galaxies on which the indicator has been calibrated, we get results in agreement with previous studies ($SFR_{MW}\sim 0.71\pm0.13\;{\rm M_\odot/yr}$) and the two methods provide consistent results. This means that, when applied to areas with characteristics similar to the extragalatic star forming regions used to calibrate it, the IR estimator gives reliable results even in the MW. On the contrary, the SFR from the source counting method is not meant to be extrapolated because is based on the local population of YSOs, so it would provide very different results depending on the observed area. 

Both methods are dominated by model errors which are not quantifiable. The evolutionary star model, the IR indicator and the Galactic extrapolation are subject to assumptions which might be negligible on averages of large numbers but might be significant when a limited sample is available. 
{ Therefore it will be important to expand the analysis described in this work to the whole Hi-GAL survey, which will cover 
the entire Galactic Plane and, with it, the vast majority of the star formation complexes. The use of a significantly larger sample of YSOs distributed across the Galaxy will also 
allow us to derive an effective Galactic SFR indicator.}

\section{Acknowledgments}

The authors acknowledge an anonymous referee for helpful comments. The activity of DE, MP and ES has been possible thanks to generous support from the Italian Space Agency via contract I/038/080/0.

\bibliographystyle{aa}

\begin{thebibliography}{}
\bibitem[Bally et al.(1999)]{Bally10} Bally, J., Anderson, L., D., Battersby, C. et al.
2010, A$\&$A, 518, L90
\bibitem[Battersby et al.(2011)]{Battersby11} Battersby, C., Bally, J., Ginsburg, A., et al.
2011, A$\&$A, 535, A128
\bibitem[Benjamin et al.(2003)]{Benjamin03} Benjamin, R., A., et al.,
2003, PASP, 115, 953
\bibitem[Benjamin et al.(2005)]{Benjamin05} Benjamin, R., A., Churchwell, E., et al.
2005, ApJ, 630, L149
\bibitem[Billot et al.(2010)]{Billot10} Billot, N., Noriega-Crespo, A., et al.
2010, ApJ, 712, 797
\bibitem[Calzetti et al.(2007)]{Calzetti07} Calzetti, D., Kennicutt, R., C., Engelbracht, C. W., et al.
2007, ApJ, 666, 870
\bibitem[Carey et al.(2009)]{Carey09} Carey, S. J., Noriega-Crespo, A., Mizuno, D. R, et al.  
2009, PASP, 121, 76
\bibitem[Chomiuk et al.(2012)]{Chomiuk12} Chomiuk, L., Povich, M., S.
2012, AJ, 142, 197
\bibitem[Covey et al.(2010)]{Covey10} Covey, K., R., Lada, C., J., et al.
2010, ApJ, 722, 971
\bibitem[Dame et al.(2001)]{Dame01} Dame, T., M., Hartmann, D., et al., 
2001, ApJ, 547, 792 
\bibitem[Dame $\&$ Thaddeus(2011)]{Dame11} Dame, T., M. $\&$ Thaddeus, P.,
2011, ApJ, 734, L24
\bibitem[Dunham et al.(2008)]{Dunham08}Dunham, M., M., Crapsi, A., Evans, II, N., J., et al.
2008, ApJ, 179, 249
\bibitem[Elia et al.(2010)]{Elia10} Elia, D., Schisano, E., Molinari, S., et al. 
2010, A$\&$A, 518, L97
\bibitem[Evans et al.(2009)]{Evans09} Evans, II, N., J., Dunham, M., M., et al.
2009, ApJS, 181, 321
\bibitem[Faimali et al.(2012)]{Faimali12} Faimali, A., Thompson, M., A., Hindson, L. et al.
2012, MNRAS, accepted
\bibitem[Giannini et al.(2012)]{Giannini12} Giannini,T., Elia, D., Lorenzetti, D., et al.
2012, A$\&$A,539, A156
\bibitem[Griffin et al.(2010)]{Griffin10} Griffin, M. J., Abergel, A., Abreu, A., et al.
2010, A$\&$A, 518, L3
\bibitem[G$\ddot{\mathrm u}$sten $\&$ Mezger(1982)]{Gusten82} G$\ddot{\mathrm u}$sten, R. $\&$ Mezger, P., G.,
1982, Vistas. Astron., 26, 159
\bibitem[Kauffmann et al.(2010)]{Kauffmann10} Kauffmann, J., Pillai, T., Shetty, R., et al. 
2010, ApJ, 716, 433
\bibitem[Kennicutt(1998)]{Kennicutt98} Kennicutt, R., C.
1998, AR$\&$A, 36, 189
\bibitem[Kennicutt et al.(2007)]{Kennicutt07} Kennicutt, R., C., Calzetti, D., Walter, F., et al.
2007, ApJ, 671, 333
\bibitem[Kroupa(2001)]{Kroupa01} Kennicutt, R., C.
2001, MNRAS, 322, 231
\bibitem[Lada et al.(2010)]{Lada10} Lada, C., J., Lombardi, M., Alves, J., F.
2010, ApJ, 724, 687
\bibitem[Lawton et al.(2010)]{Lawton10} Lawton, B., Gordon, K., D.,  Babler, B., et al.
2010, ApJ, 716, 453
\bibitem[Lewis $\&$ Bridle(2002)]{Lewis02} Lewis, A., $\&$ Bridle, S.
2002, Phys. Rev., 66, 103511
\bibitem[Li et al.(2010)]{Li10} Li, Y., Calzetti, D., Kennicutt, R., et al.
2010, ApJ, 725, 667
\bibitem[Martinavarro et al.(2012)]{Martinavarro12} Martinavarro, S., et al.
2012, in preparation
\bibitem[McKee et al.(2003)]{McKee03} McKee, C., F., $\&$ Tan, J., C., et al.
2003, ApJ, 585, 850
\bibitem[Molinari et al.(2008)]{Molinari08} Molinari, S., Pezzuto, S., R., et al.
2008, A$\&$A, 481, 345
\bibitem[Molinari et al.(2010a)]{Molinari10a} Molinari, S., Swinyard, B., et al.
2010, PASP, 122, 314
\bibitem[Molinari et al.(2010b)]{Molinari10b} Molinari, S., Swinyard, B., et al.
2010, A$\&$A, 518, L100
\bibitem[Molinari et al.(2011)]{Molinari11} Molinari, S., Schisano, E., et al.
2010, A$\&$A, 530, A133
\bibitem[Murray $\&$ Rahman(2010)]{Murray10} Murray, N., Rahman, M.,
2010, ApJ, 709, 424
\bibitem[Paladini et al.(2004)]{Paladini04} Paladini, R., Davies, R., D., et al.
2004, MNRAS, 347, 237
\bibitem[Paradis et al.(2010)]{Paradis10} Paradis, D., Veneziani, M., Noriega-Crespo, A., et al.
2010, A$\&$A, 520, L8
\bibitem[Pilbratt et al.(2010)]{Pilbratt10} Pilbratt, G., Riedinger, J. R., et al.
2010, A$\&$A, 518, L1
\bibitem[Poglitsch et al.(2010)]{Poglitsch10}Poglitsch, A.,  et al.
2010, A$\&$A, 518, L2
\bibitem[Preibisch et al.(1993)]{Preibisch93} Preibisch, T., Ossenkopf, V., et al.
1993, A$\&$A, 279, 577
\bibitem[Reed(2005)]{Reed05} Reed, B., C., 
2005, AJ, 130, 1652
\bibitem[Robitaille et al.(2006)]{Robitaille06} Robitaille, T., P., Whitney, B., A., et al. 
2006, ApJS, 167, 256
\bibitem[Robitaille et al.(2008)]{Robitaille08} Robitaille, T., P., Meade, M., R., et al. 
2008, AJ, 136, 2413
\bibitem[Robitaille $\&$ Whitney(2010)]{Robitaille10} Robitaille, T., P., Whitney, B., A. 
2010, ApJ, 710, L11
\bibitem[Russeil(2003)]{Russeil03}Russeil, D.
2003, A$\&$A, 397, 133
\bibitem[Russeil et al.(2011)]{Russeil11}Russeil, D., Pestalozzi, M., Mottram, J., C., et al. 
2011, A$\&$A, 526, A151
\bibitem[Saraceno et al.(1996)]{Saraceno96} Saraceno, P., Andre, Ph., Ceccarelli, C., et al.
1996, A$\&$A, 309, 827
\bibitem[Tibbs et al.(2012)]{Tibbs12} Tibbs, C., Paladini, R., Compiegne, M. et al.
2012, ApJ, 754, 94
\bibitem[Traficante et al.(2011)]{Traficante11} Traficante, A., Calzoletti, L., Veneziani, M. et al.
2011, MNRAS, 416, 2932


\end{thebibliography}

\end{document}